# Laser Acceleration toward PeV Feeling the Texture of Vacuum

T. Tajima*

Faculty of Physics, Ludwig Maximilian University of Munich, Garching, 85748, Germany,

M. Kando

Japan Atomic Energy Agency, 8-1-7 Umemidai, Kizugawa, Kyoto, 619-0215, Japan

and

M. Teshima

Max Planck Institute for Physics, Munich, 80805, Germany

*: Blaise Pascal Chair, Ecole Normale Superieure

PACS: 41.75.Jv , 98.70.Rz, 03.30.+p

**Abstract**

Identified is a set of ballpark parameters for laser, plasma, and accelerator technologies that are defined for accelerated electron energies reaching as high as PeV. These parameters are carved out from the scaling laws that govern the physics of laser acceleration, theoretically suggested and experimentally explored over a wide range in the recent years. We extrapolate this knowledge toward PeV energies. In the density regime on the order of $10^{16}$ cm$^{-3}$, it is possible to consider the application of the existing NIF (or LMJ) or its extended lasers to their appropriate retrofitting for this purpose. Although the ambition of luminosity is not pursued, such energies by themselves may allow us to begin to feel and study the physics of the 'texture of vacuum'. This is an example of fundamental physics exploration without the need of luminosity paradigm. By converting accelerated electrons with extreme energies to like energy gamma photons, and let them propagate through vacuum over a sufficient distance, these extremely high energy (and therefore short wavelength) photons experience smallest vacuum structures and fluctuations. If we can measure the arrival time differential and thus the gamma photon speed as a function of different energies such as 0.1 PeV vs 1 PeV, say within attoseconds accuracy, we can collect valuable data if and how gamma photons still obeys the premise of relativity or the vacuum texture begins to alter such fundamentals. The only method currently available to look at this problem may be to study astrophysical data of the primordial gamma ray bursts (GRBs), which are compared with the presently suggested approach.

**1. Introduction**

Since the laser based particle acceleration was conceived [1], energies of laser-accelerated electrons have increased with the advance of laser technologies and better control and understanding of the experiments [2-14]. Leemans et al. reported GeV laser acceleration of electrons using 40 TW, 40 fs laser pulses with a 3 cm plasma channel [2]. With these and other past experiments it is now evident that the laser accelerated electron energies scale inversely proportional to the plasma density, as predicted by Tajima and Dawson [1]. An experimental data summary is shown in Fig. 1. This figure shows this tendency of energy gain as a function of the plasma density $n_e$: Unmistakably, the energy gain rises continually and linearly in the plot, as the electron density falls. This presents us opportunities to consider a design of experiments toward 10 GeV, 100 GeV, and 1 TeV energies based on laser wakefield acceleration, including the application to a future collider, following the simple and yet robust scaling law [15-22].

First, in the present paper, however, we do not pursue high repetition rates and high luminous experiments as required by colliders. This exempts and relieves us from the constraints typical of colliders that require lasers (or for that matter, any driver of accelerator) to be highly efficient and with high average fluence. The physics we try to reach out is confined to single-shot or low repetition experiments, albeit with extreme high energies (much beyond TeV now unconstrained by the collider physics requirements [15-17][23]). Second, in order to preserve the laser energy only for the purpose to excite the laser wakefield, we should avoid the wavebreaking, which leads to unnecessary electrons to be trapped and wakefield energies diverted to these particles. We thus need to operate in or near the one-dimensional wakefield. It is well known that the 1D wakefield is most robust, while 2D/3D wakefields lead to much easier wavebreaking and trapping of unwanted electrons. The choice of near 1D wakefield operation also aids us to be in a relatively simple acceleration control. Third, on the other hand, in order to reach highest energies within shortest possible distances, we ought to optimize physical parameters to the limit allowable. For example, we need to maximize the accelerating gradient within the limit of 1D-like wakefield with smallest allowable radius of laser focus, which minimizes the needed laser energy. In order to increase the accelerating gradient, we enter in the nonlinear wakefield regime (unlike the typical requirement of the linear wakefield regime for the collider operation, where the extreme low emittance preservation is necessary in order to realize the high luminosity at the tight focused collision point). At this high intensity nonlinear wakefield regime it is likely for the laser power to exceed the self-focusing threshold. By carefully choosing the laser radius at the equilibrium radius of the self-focused beam, however, we should be optimizing both the highest possible fields and the near 1D wakefields.

Professor Atsuto Suzuki [24] has challenged us if we can come up energies of even PeV. This was first toyed by E. Fermi [25], whose vision goes to girdle the entire earth by the circular accelerator to reach this goal. We try to see if we can meet Suzuki's visionary challenge of PeV with our vision of laser acceleration. In what follows we try to identify ballpark parameters required for PeV electron acceleration based on the laser wakefield acceleration process [1][15-18] . We find the required laser parameters do not completely go mismatched with the present or near-term lasers and their extensions. In these extreme energies the physics we can explore may not be those extended from the present day colliders, but it rather poses a different type of experiments. We present an application of this accelerator as a possible sample illustration.

**2. Scaling laws of laser wakefield acceleration**

Consider the possibility of acceleration to reach PeV energies via the laser wakefield acceleration. In order to reach the highest energy possible with the judicious deployment of the currently available laser or a near future prospect of lasers, our approach is as follows. We thus adopt one-dimensional, non-linear wakefield acceleration which is operated nearly at its limits: the laser field is in the neighborhood of the limit of one-dimensional wavebreaking and the laser spot radius is close to the limit of one-dimensional wake structure. We notice that this adoption is conceptual at this moment, and detailed calculations are required for the design of a practical accelerator. However, as we see in Fig. 1, which shows the experimental achievements so far, the one-dimensional approximation adopted here does not fail to predict the energy gains of experimental data. Two orders of magnitude difference between the theory and experiments might come from the condition that the laser irradiances at some experiments were not high. Indeed, the aimed goal is far from the currently obtained experimental domains in terms of energy gain. The blank region is left to be filled in as a future task in the laser wakefield acceleration research.

Summarized here are the scaling laws of laser wakefield acceleration that fit for the above approach and have been theoretically presented and experimentally observed in the past works [18][21]. If we take the scaling law based on a one-dimensional, nonlinear theory of the wakefield acceleration [21], the energy gain $\Delta E$ of electrons per stage in a highly nonlinear regime ($a_0 \gg 1$) is approximately expressed as

$$\Delta E \approx \gamma_{ph}^2 a_0^2 m_0 c^2, \qquad (1)$$

where $m_0$ is the electron rest mass, $c$ is the speed of light, $a_0 = eE_L/m_0\omega_0 c$ is the normalized vector potential of the pump laser with the electric field of $E_L$ and the frequency of $\omega_0$, $\gamma_{ph} = \left(1 - v_{ph}^2/c^2\right)^{-1/2}$ and $v_{ph}$ is the phase velocity of the wakefield, $n_{cr}$ and $n_e$ are the critical density and the plasma density, respectively. Since we consider the laser wakefield acceleration in the nonlinear regime, the laser amplitude is maximized to be close to the wave-breaking limit but not reaching or exceeding it in order to avoid deleterious wave-breaking effects. This condition requires the normalized laser amplitude to satisfy $a_0 \leq 2\gamma_{ph}^{1/2}$. On the other hand, in a relatively strong drive, the wave assumes a steep profile and thus once again nearly one-dimensional physics may become important in the immediate vicinity of this sharp gradient. In fact Koga et al.'s simulation [26] saw a steep wave gradient and much flattened wave front even though his laser pulse was relatively narrow: The frontal part of the wave is appropriate for accelerating positrons [27] (or other positively charged particles), while the rear part for electrons (not all the parameters in Ref. [26] scale with what we suggest here).

The acceleration length is limited by the dephasing length or the pump depletion length. The dephasing and pump depletion lengths may be given by [15]

$$L_d = \gamma_{ph}^2 \lambda_p \frac{a_0}{\pi\sqrt{2}} \qquad (2)$$

$$L_{pd} = \gamma_{ph}^2 \lambda_p \frac{\sqrt{2}a_0}{\pi} \qquad (3)$$

where $\lambda_p = 2\pi c/\omega_p$ is the linear and nonrelativistic plasma wavelength. The optimum condition for laser wakefield excitation is realized [15] when the laser pulse length at full width half maximum (FWHM) $c\tau$ matches this plasma wavelength, *i.e.*

$$c\tau \approx \frac{\sqrt{2\ln 2}}{\pi} \approx 0.37\lambda_p. \qquad (4)$$

The peak power $P$ of the laser is $P = I\pi w_0^2$, where $I$ is the peak irradiance of the laser. The total laser energy $E_L = P\tau$ necessary in the quasi-one dimensional case is expressed as

$$E_L[\text{J}] = \left(\frac{a_0}{0.86\lambda_0[\mu m]}\right)^2 \times \frac{\pi(w_0[\mu m])^2}{100}\tau[\text{ps}]. \qquad (5)$$

We notice that the self-focusing condition $P > P_c = 2c(mc^2/e)^2(\omega/\omega_p)^2$ is always satisfied when $a_0 > 2^{3/2}/\pi \sim 0.9$ ($w_0 = \lambda_p$). After the self-focusing, the spot radius of the laser becomes $\sim (a_0)^{1/2} c/\omega_p$, which is similar to our assumption (To make more one-dimensional, one may take a larger spot radius to avoid self-focusing by adopting a defocusing waveguide, which has a lower refractive index in the radial center as opposed to the usual uniform waveguide).

As a comparison, we list the scaling when three-dimensional effects are important (e. g. $w_0 \leq c/\omega_p$) the energy obtained by laser acceleration may become slightly more complicated. Consider the case when the laser pulse is intense enough to make a cavity behind the laser pulse, i.e. $a_0 > e\varphi/(mc^2)$, where $\varphi \approx 4\pi n_e e w_0^2$ is the electrostatic potential of the wake. According to the study [28], in this case we obtain

$$\Delta E \approx \frac{\pi^2 m_0 c^2 w_0^4}{8\lambda_0^2 \lambda_p^2}, \qquad (6)$$

where the laser spot size $w_0$ is related to

$$w_0 \approx \frac{1}{\pi}\gamma_{ph}\lambda_0 \qquad (7)$$

and the cavity longitudinal size is of the order of the transverse size, $w_0$. In this tightly focused case, optical guiding is required to extend the acceleration length. According to Ref. [29], the matched spot size $w_M$ in the capillary, with the radial plasma density profile $n_e(r) = n_0 + \Delta n_e (r/R)^2$, is given by

$$w_M = \left( \frac{R^2}{\pi r_e \Delta n_e} \right)^{1/4}, \tag{8}$$

where $R$ is the radius of the capillary wall. If we set $w_0 = c/\omega_p = \lambda_p/2\pi$ in order to avoid self-focusing or filamentation of the pump pulse, the energy gain scales as $\Delta E \propto 1/n_e$.

## 3. Ballpark parameters of laser electron accelerator toward PeV

The scaling law dictates some three orders of magnitude density reduction from most current experimental parameters with the typical density at $10^{18}$ cm$^{-3}$ in order to carry out experiments in the range toward energies of PeV in a single stage. This in turn allows us to extend the laser pulse length by an order of magnitude, to typically on the order of ps, instead of tens of fs. In a multi-stage approach, say $10^2$-$10^3$ stages, in order to reach these energies, the density is higher and pulse length shorter. The preferred laser technology of recent laser acceleration experiments has been that of Ti:sapphire because of its large frequency bandwidth, but for longer pulses a wider range of lasers become permissible.

Here we take a few typical numerical examples at various initial laser intensities as listed in Table 1 for the PeV energy acceleration. We assume that the laser wavelength is 1 μm and the spot size of the laser is $w_0 \approx \lambda_p$ to make the operation in the 1D regime. We range the number of stages of laser wakefield acceleration. According to Eq. (1), the required plasma density is calculated and thus other parameters including the laser intensity, or the normalized vector potential $a_0$, are automatically determined. The number of electrons is calculated based on the formula by Katsouleas et al.[30]. The total energy gain is just multiplying the single-stage energy gain by the number of stages. If we take the number of stages, $N_{stage}$= 1000, we need $n_e = 1.8 \times 10^{17}$ cm$^{-3}$ to reach 1 PeV total energy gain. Under the current choice of ballpark we suggest that the optimum laser parameters are 4.1 MJ, 42 PW, and 0.098 ps. This is the case studied in Table I case III. The acceleration length per stage is ~2 m and the total acceleration length is 2 km. The total acceleration length here means the sum of the individual stages without including the necessary matching sections, i.e.

focusing optics of the electron beam and driving laser. The usual electromagnetic focusing system for electron beams may require substantially longer matching sections. However, it may be possible that the adoption of plasma lens lowers the length to an affordable size. In this choice the required laser pulse may strain the existing laser technology, as we shall discuss below. To ameliorate such a situation, the introduction of the nonuniform plasma density profile with a density initially lower than the value taken here might bring in some room to maneuver: the laser pulse compression may take place through the nonlinear interaction with the plasma [31] to fit more adequately and gradually increase the density to the value considered here.

**4. Possible experiment and its ramification in comparison with astrophysical data**

Ellis et al [32, 33] (note: a jump in the refs numbers) have suggested that the quantum mechanical fluctuations with wavelengths on the order of quantum gravity origin may amount to the effective slowdown of the photon velocity, if the energy of the photon is high enough and its wavelength short enough to see such scale lengths of fluctuations. These fluctuations may be directly tied with the length scale inverse of the Planck mass or may be even longer. There are other theories [34,35] that suggest that the photon velocity varies when its energy goes up. Of course, it is of immense importance to examine if such phenomenon appears at all and if such theories are correct (if any) or when such phenomenon begins to manifest. This is a fundamental test of the special theory of relativity and perhaps a prelude to a glimpse into quantum gravity. We envisage that such a test can be one of the candidate experiments that need not demand the high luminosity that a collider would. Thus we wish to consider this sample experiment in some more detail in this section.

At this moment, however, barring our PeV candidate experiment, all we can do is left to astrophysical observations to ask such questions. This is in part due to the fact that it is believed that if such a phenomenon exists at all, it should be so high an energy that is simply much beyond the reach of the present day accelerator on earth. On the other hand, we are learning a lot recently about the high energy gamma ray emission from very fast flares from Active Galactic Nuclei [36, 37] and Gamma Ray Bursts (GRBs) which are known as brightest astrophysical objects [38, 39]. The energy dependence of light velocity has been tested using photon beams from such objects.

GRBs are categorized into two types, long one and short one. It is generally believed that long and short GRBs may be related to the supernova / hypernova collapse, and to the merger of two neutron stars (or some other very compact stars), respectively. In both GRBs there are two components in gamma rays. One is the component described with the band function which ranges between 30 keV and 10 MeV and can be described by two power-law spectra before and after the peak energy around 300 keV; the other is the extra delayed component ranging between 30 keV and 30 GeV (or beyond) and can be described with a simple power-law without a cut-off and break [38, 39]. These two components are believed to arise from different origins / from different emission regions of GRBs.

Since GRBs are the brightest astrophysical objects and with a short characteristic time envelope, they can serve as an ideal searchlight to explore the deepest Universe. The primordial GRBs have been thus cherished to look for their time history of their arrival to the Earth observatories over nearly an entire length of the Universe. If there is any energy dependence in the photon velocity, the larger energy photon would arrive later than the less energy ones in the give GRB. Many of GRBs studies so far [38, 39], in fact, show this tendency. Furthermore, this tendency seems consistent with each other; in another word, most of these observations show a similar arrival differential as a function of the energy of gamma. Except for the fact that it appears that the latest short GRB observed by the Fermi Observatory [40], which may be showing a less differential time arrival, though it too shows that the higher the energy of gammas, the later they arrive.

On the other hand, one may argue that the delayed arrival of higher energy gammas is not due to the propagation property in the space between the GRB and the Earth, but rather the reflection of the genesis of GRBs and their mechanism of the particle acceleration to high energies at the time of the burst (e.g., [41]). One might argue that the higher electron energies are, the longer time it takes to get accelerated and thus the emission of gammas with higher energies should appear later. If this is the case, what we are observing is simply the property of GRB and its acceleration mechanism of high energy electrons in the GRB jet. It is not easy to dismiss such an argument when we wish to refer to the property of vacuum for the gamma ray propagation. We would be left to speculate which is more likely at this time.

Thus it would be scientifically valuable to be able to have a controlled terrestrial experiment that can be determine the gamma speed as a function of its energy that is not depending upon the genesis of that gamma beam, as suggested in Secs 2-3. This may become possible if our accelerated electron reaches as high energies as PeV. Consider the following experimental scenario. The energies of the highest energy gammas from GRB are typically GeV, while the cosmic distance is on the order of $10^{28}$ cm. If we take the length of our vacuum tube is about a km, the time differential we need to ascertain is on the order of sub fs, in order to meet or discriminate against the GRB observations of a second to 10s of seconds. We understand that it takes ingenious experimental innovations unexplored so far to measure the arrival time of two gamma photons (or beams of photons) with two different energies, say PeV and 0.1 PeV with ultrafast accuracy. No one seems to have ever looked at PeV gamma arrival detection in such a time differential regime and this remains a challenge.

We have not started systematic experimental research of how to detect ultrahigh energy gamma particles and differentiate the arrival time with ultra high time resolution. However, we venture at least some attempt into a possible detection technique development here. It has been pointed out by Narozhny some 40 years ago [42] (more recently [43]) that an ultrahigh energy gamma-particle can assist to break down the vacuum with substantially suppressed threshold electric field compared with the well-known Schwinger value. This is the nonlinear QED effect. The probability of the vacuum breakdown is derived as

$$P(E) \propto \exp\left[-\frac{8}{3}\left(\frac{E_s}{E}\right)\left(\frac{mc^2}{\hbar\omega}\right)\right] \tag{9}$$

where $E_s$ the Schwinger field, $\hbar\omega$ is the gamma energy, $E$ is the applied electric field in vacuum such as a laser. With a PeV gamma-ray particle, the exponent factor of (9) is reduced by the ratio of MeV to PeV ($mc^2/\hbar\omega$) over the expression of Schwinger's without the presence of a gamma particle. This means that the vacuum breakdown field plummets from the value of $10^{16}$ V/cm to $10^{10}$ V/cm.

We suggest that by employing time-synchronized somewhat intense laser field (at $10^{10}$ W/cm$^2$) at the "goal line" of the gamma-photon arrival, we cause sudden breakdown of vacuum and its avalanched particles of e⁻e⁺ as soon as one of the high energy gamma particles arrives. The PeV gamma particle facilitates to trigger the vacuum breakdown. The time scale of breakdown is far faster than fs. The exploitation of this phenomenon should allow an ultrafast signal of the PeV gamma-photon arrival. Since the trigger phenomenon is exponentially sensitive, we could play a game of adjusting the value of the laser field to see and differentiate different types of trigger phenomenology and parameters, depending upon the gamma particle energies.

We obviously need a lot more detailed experimental planning and developments of such an idea in the future. Further, the delay of gamma-photon arrival to the "goal line" due to the presence of low-density electrons is an important factor that determines the "noise" to our "signal". One of the noises or uncertainties about the time differential may arise from residual gas electrons in our vacuum tube in which gamma particles travel. We may be able to evaluate this time delay as follows.

The dielectric refractive index of the plasma with the density $n$ is given by

$$n = \left(1 - \frac{\omega_p^2}{\omega^2}\right)^{1/2}, \tag{10}$$

where $n$ sets the phase velocity of light as and the group velocity as $n = \omega/kc = v_{ph}/c$ and the group velocity as

$$v_{gr} = c\left(1 - \frac{\omega_p^2}{\omega^2}\right)^{1/2}. \tag{11}$$

The difference between $c$ and $v_{gr}$ is

$$\frac{c - v_{gr}}{c} \cong \frac{1}{2}\frac{\omega_p^2}{\omega^2}. \tag{12}$$

This amount is extremely small for high-energy gamma particles with PeV energies. If the gas pressure of the 'vacuum' is as low as $10^{-6}$ Pa, $(c - v_{gr})/c$ is as small as $10^{-44}$ for PeV gamma photons. On the other hand, the expected (if it ever arises or the margin we try to establish) deviation of the speed of light in extreme high energies ($\hbar\omega$) of PeV in our suggested experiment is as high as $\Delta c(\omega)/c \sim 10^{-10}$. Therefore, we seem not to be excluded from the possibility to test, feel, and detect the texture of vacuum that may arise from the quantum gravity effects and the subsequent phenomenon of the energy-dependent speed of light in such an experiment.

## 5. Discussion and conclusions

We have presented the possibility that utilizing the existing large energy lasers or its future extension, we can chart out a scientific path to reach for PeV energies by the laser acceleration. The laser wakefield acceleration (LWFA) is capable of very compact and intense acceleration far beyond the conventional accelerator approach. Reaching such energies as PeV appears only possible by such a new enabling method. We have then established a set of principles and associated parameters that allow us to reach for these energies. By adopting multi-MJ laser capability that exists in National Ignition Facility [44] (and soon completing Laser Mega Joule) and other future outgrowth of these lasers, we employ the (approximately) 1D and strongly nonlinear regime of LWFA to optimize the beam quality, the accelerating gradient, and other physical attributes. Based on this approach and the scalings known from the past theoretical and experimental investigations, we are led to show that there exist a set (or sets) of parameters that allow us to envision a PeV accelerator.

These ideas and parameters are of a fundamental principle of this acceleration method and not necessarily scrutinized for engineering details. Thus in the future we need to look for more in-depth studies and experimental investigations to ascertain the possibility for realizing such extreme energies using the LWFA. Nonetheless, it is very encouraging that already today's laser technology is at or near the ballpark of the necessary requirement as to the laser energy is concerned. No doubt that we need to learn plenty more on how to accomplish PeV acceleration using this method in the future.

Even though it appears to us not possible to make a PeV accelerator into a collider, because of its too severe requirements for luminosity, we wish to seek other applications at the energy frontier. We have suggested at least one such a candidate. If we use PeV electrons to produce PeV photons (gamma particles), these photons serve us to investigate new physics. We have suggested that with energy varying gamma particles, we can measure the arrival time differentiation of these gamma particles over some distance, say a km at or around PeV. According to some theories on quantum gravity and other alternative theories, the Lorentz transformation with respect to the speed (or the Lorentz factor $\gamma$) is no more invariant, but rather dependent on energies of the gamma photons. According to some of these theories, it is possible that when the energies of photons become as large as PeV, such effects may be magnified so much to become observable. This is precisely what we suggest here in this article.

So far it appears that the only way to test such possibilities and theories is through astrophysical observations. Thus astrophysicists have ventured to use primordial Gamma Ray Bursts (GRBs) to observe their arrival differential depending on their energies (frequencies). GRBs are the brightest objects in the Universe and thus we should be able to detect the most ancient and thus farthest. In fact the primordial GRBs can make us encompass the entire distance of the cosmos, thereby enabling us to magnify the time differential at maximum. Thereby, astrophysicists, amazingly, seem to have seen some time differentials of gamma particle arrivals from GRBs with statistical significant amounts. These indicate, by and large, the more energetic gamma photons are, the later they see to arrive, in crude agreement with what these quantum gravity theories would predict. However, there remains a large body of discourses as to the nature of these time delays. For example, the time delays may be due to the GRBs source characteristics: the higher the gamma particle energy is, the longer it takes for these particles to get accelerated, thus such time differential, but not due to the vacuum property of the photon propagation sped over the Universe distance. Also there seems to have some statistical debates among various observations to date. These are the nature of the astrophysical observations and cannot be easily eradicated. It is thus ideal if we can come up with controlled experiments. This is what our PeV acceleration should be able to meet. It may be that this would pose the severest terrestrial test of Einstein's Special Theory of Relativity ever.

We have begun to explore an ultrafast optical detection method of the gamma particle arrival differential. This seems not out of bound of physical reality. Although it provides so far only a crude principle to test such grandiose effects, to the first order it seems that we have not encountered fundamental difficulties. Of course, more details of such ideas and methods need to be studied. In addition, we could imagine more applications of PeV electrons (or other particles such ions) at or near PeVs. We look for more investigations in this direction in the future. Finally, as to ion acceleration in this PeV LWFA, except for the first few GeV booster / injector, ion acceleration is not so much different from electron acceleration in this linear accelerator. It might have some potential benefits for less stringent orbital requirements, such as the benefit of the lack of betatron radiations. If one has tangible experimental incentives for the PeV hadron sector physics, it would be of interest to pursue this avenue as well.

Acknowledgements:

We would like to thank the encouragements, camaraderie, and advices of A. Suzuki, G. Mourou, the late Y. Takahashi, D. Habs, E. Esarey, W. Leemans, H. Sato, D. Jaroszynski, M. Downer, S. Cheshkov, S. Bulanov, T. Esirkepov, J. Koga, F. Takasaki, J. Urakawa, K. Nakajima, C. Barty, E. Moses, C. Labaune, and D. Normand. This work was partially supported by KAKENHI (No. 21604008).

We wish to dedicate this paper to the late Professor Yoshiyuki Takahashi of Huntsville, Alabama, who taught us so much of extreme high energy phenomena and GRBs.

**References**

[1] T. Tajima and J. M. Dawson, Phys. Rev. Lett., **43**, 267, (1979).

[2] W. P. Leemans *et al*., Nature Physics, **2**, 696 (2006).

[3] E. Miura *et al*., Appl. Phys. Lett. **86**, 251501 (2005).

[4] S. P. D. Mangles *et al.*, Nature, **431**, 535 (2004).

[5] C. G. R. Geddes *et al*., *ibid*, 538 (2004).

[6] J. Faure *et al*., *ibid*, 541 (2004).

[7] A. Yamazaki *et al.*, Phys. Plasmas,**12**, 093101 (2004).

[8] T. Hosokai *et al.*, Phys. Rev. E, **73**, 036407 (2006).

[9] C.-T. Hsieh *et al.*, Phys. Rev. Lett. , **96**, 095001 (2006).

[10] B. Hidding *et al.*, Phys. Rev. Lett., **96**, 105004 (2006).

[11] M. Mori *et al.*, Phys. Lett. A, **356**, 146 (2006).

[12] S. P. D. Mangles *et al*., Phys. Rev. Lett., **96**, 215001 (2006).

[13] S. Kneip *et al*., Phys. Rev. Lett., **103**, 035002 (2009).

[14] D. H. Froula *et al*., Phys. Rev. Lett., **103**, 215006 (2009).

[15] E. Esarey and W. Leemans, Rev. Mod. Phys. **81**, 1229-1285 (2009).

[16] M. Xie, T. Tajima, K. Yokoya, and S. Chattopadhyay, in *Proceedings of the Advanced Accelerator Concepts Workshop, Lake Tahoe, 1996*, edited by S. Chattopadhyay, J. McCullough, and P. Dahl (AIP Conf. Proc. 398, New York), pp.

233-242.

[17] S. Cheshkov, T. Tajima, W. Horton, and K. Yokoya, Phys. Rev. ST Accel. Beams **3**, 071301 (2000).

[18] E. Esarey and P. Sprangle, IEEE Trans. on Plasma Science, **24**, 252 (1996).

[19] S. V. Bulanov, F. Pegoraro, A. M. Pukhov, A. S. Sakharov, Phys. Rev. Lett. **78**, 4205 (1997).

[20] M. Kando et al., in *Proceedings of Laser-Driven Relativistic Plasmas Applied for Science, Industry, and Medicine, Kyoto, 2007,* edited by S. V. Bulanov and H. Daido (AIP Conf. Proc. 1024, New York) pp. 197-207.

[21] E. Esarey and M. Pilloff, Phys. Plasmas **2**, 1432 (1995).

[22] W. Lu et al., Phys. Rev. Lett., **96**, 165002 (2006).

[23] T. Tajima, and G. Mourou, Phys. Rev. STAB **5,** 031301(2002).

[24] A. Suzuki, Outlook: Accelerator Science (ICFA, SLAC, 2008), http://www-conf.slac.stanford.edu/icfa2008/Suzuki_103108.pdf

[25] E. Fermi, PeV; http://en.wikipedia.org/wiki/List_of_accelerators_in_particle_physics

[26] J. Koga, K. Nakajima, and K. Nakagawa, in *Superstrong Fields in Plasma*, ed. M. Lontano et al. (AIP, New York, 2002), p.126.

[27] T. Zh. Esirkepov et al., Phys. Rev. Lett., **96,** 014803 (2006).

[28] S. V. Bulanov, Plasma Phys. Control. Fusion **48**, B29 (2006).

[29] N. A. Bobrova et al., Phys. Rev. E **65**, 016407 (2002).

[30] T. Katsouleas, S. Wilks, P. Chen, J. M. Dawson, and J. J. Su, Part. Accel. **22**, 81 (1987).

[31] J. Faure et al., Phys. Rev. Lett., **95**, 205003 (2005).

[32] J. Ellis, N. Mavromatos, and D. Nanopoulos, Phys. Lett. B **665**, 412 (2008).

[33] G. Amelino-Camelia, J. Ellis, N. Mavromatos, D. Nanopoulos, and S. Sarkar, Nature **393**, 763 (1998).

[34] H. Sato and T. Tati, Prog. Theor. Phys. **47**, 1788 (1972); H. Sato, arXiv:astro-ph/0005218v1 10May2000.

[35] S. Coleman and S. Glashow, Phys. Lett. B **405**, 249 (1997); Phys. Rev. D **59**, 116008 (1999).

[36] MAGIC Collaboration, Phys. Lett. B **668**, 253 (2008).

[37] F. Aharonian et al., Phys. Rev. Lett. **101**, 170402 (2008).

[38] M. M. Gonzalez *et al.*, Nature (London) **424**, 749 (2003).

[39] A. A. Abdo et al. Astrophys. J., **706**, L138 (2009).

[40] A. A. Abdo *et al.*, Nature (London) **462**, 331 (2009).

[41] Y. Takahashi, L. Hillman, and T. Tajima, in <u>*High Field Science*</u>, Eds., T. Tajima, K. Mima, and H. Baldis (Kluwer, NY, 2000) p.171.

[42] N. B. Narozhny, Sov. Phys. JETP **27**, 360 (1968).

[43] V. N. Baier and V. M. Katkov, arXiv:0912.5250v1 [hep-ph] 29 Dec 2009.

[44] G. Miller, E. Moses, and C. Wuest, Opt. Eng. **43**, 2841 (2004); C. A. Haynam et al., Appl. Opt. **46**, 3276 (2007).

**Figure and table captions**

Figure 1. Electron energy as a function of the plasma density observed in experiments [2-14]. The solid line shows the fitted curve of $\Delta E/mc^2 = 2\times(1.6\times10^{19}/n_e)^{3/2}$ gleaned and $\chi^2$-matched from all these experimental data, and the broken line shows the theoretical scaling ($\Delta E/mc^2 = 2\times(1.7\times10^{21}/n_e)^{3/2}$).

Table 1. Sample parameters for PeV proof-of-principle laser acceleration of electrons and positrons. Case III staging makes the required laser energy come within a parameter domain reachable with the latest laser technology similar to that of NIF or LMJ .

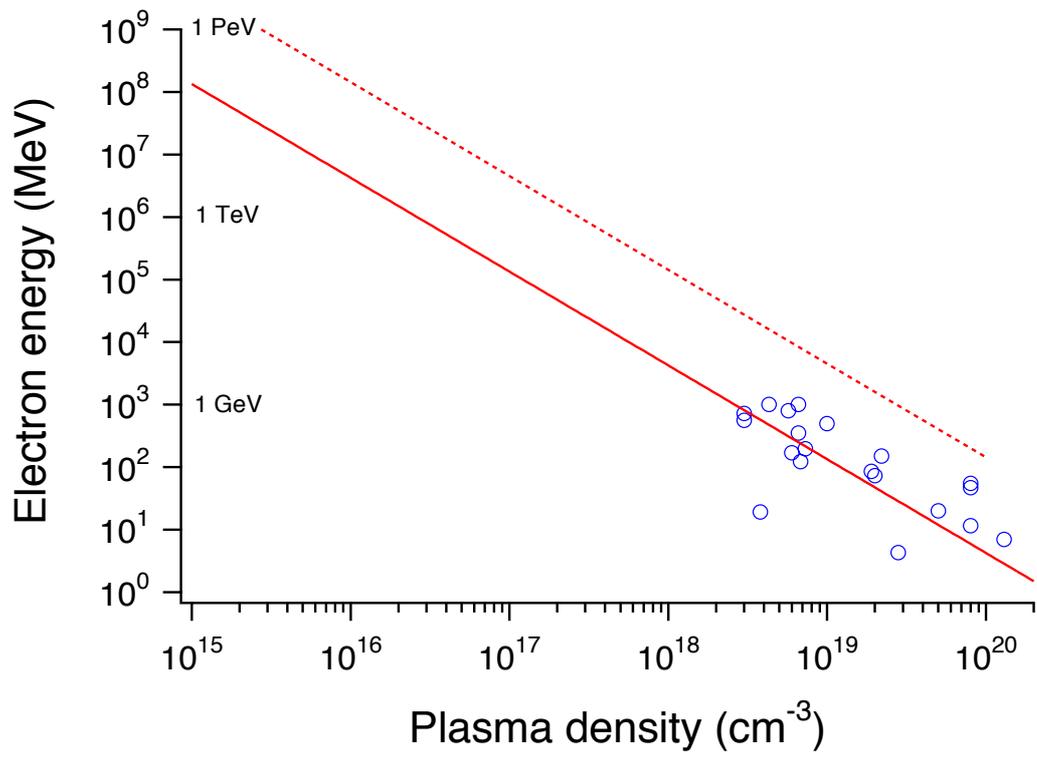

Figure. 1

Table 1 .

| Parameters | Symbol | Case I | Case II | Case III | unit |
|---|---|---|---|---|---|
| total energy gain | $\Delta W$ | 1 | 1 | 1 | PeV |
| total laser energy | $E_{L,t}$ | $4.1 \times 10^2$ | 8.8 | 4.1 | MJ |
| number of stages | $N_{stage}$ | 1 | 100 | 1000 | |
| plasma density | $n_e$ | $1.8 \times 10^{15}$ | $3.9 \times 10^{16}$ | $1.8 \times 10^{17}$ | cm$^{-3}$ |
| gamma factor | $\gamma_{ph}$ | $7.9 \times 10^2$ | $1.7 \times 10^2$ | 79 | |
| wavelength | $\lambda$ | 1 | 1 | 1 | um |
| norm. laser amplitude | $a_0$ | 56 | 26 | 18 | |
| laser energy/stage | $E_{L,l}$ | $4.1 \times 10^4$ | 88 | 4.1 | kJ |
| peak power | $P$ | $4.2 \times 10^4$ | $4.2 \times 10^2$ | 42 | PW |
| pulse duration | $\tau$ | $9.8 \times 10^2$ | $2.1 \times 10^2$ | 98 | fs |
| pump depletion length | $L_p$ | $1.2 \times 10^4$ | 57 | 3.9 | m |
| dephasing length | $L_d$ | $6.2 \times 10^3$ | 29 | 2 | m |
| total acc length | $L_{acc,t}$ | $6.2 \times 10^3$ | $2.9 \times 10^3$ | $2.0 \times 10^3$ | m |
| spot radius | $w_0$ | $7.9 \times 10^2$ | $1.7 \times 10^2$ | 79 | μm |
| number of electrons | $N_{beam}$ | 1.7E+11 | 1.7E+10 | 5.5E+09 | |